\documentclass[12pt, onecolumn,draftclsnofoot]{IEEEtran}
\usepackage[final]{graphicx}
\usepackage[final]{graphicx}
\usepackage{amsmath,mathrsfs,amssymb}
\newcommand{\bc}{\begin{center}}
\newcommand{\ec}{\end{center}}
\newcommand{\bdm}{\begin{displaymath}}
\newcommand{\edm}{\end{displaymath}}
\newcommand{\beq}{\begin{equation}}
\newcommand{\eeq}{\end{equation}}
\newcommand{\bfl}{\begin{flushleft}}
\newcommand{\efl}{\end{flushleft}}
\newcommand{\bt}{\begin{tabbing}}
\newcommand{\et}{\end{tabbing}}

\usepackage{here}
\numberwithin{equation}{section} \allowdisplaybreaks

\begin{document}

\title{Understanding Transcriptional Regulation Using De-novo Sequence Motif Discovery, Network Inference and Interactome Data}

\author{ Arvind~Rao, Alfred~O.~Hero~III, David~J.~States, James~Douglas~Engel
\thanks{Arvind~Rao and Alfred~O.~Hero,~III are with the Departments of Electrical
Engineering and Computer Science, and Bioinformatics at the
University of Michigan, Ann Arbor, MI-48109, email: [ ukarvind,
hero] @umich.edu}
\thanks{James~Douglas~Engel is with the Department of Cell and Developmental Biology at the University of Michigan, Ann Arbor, MI-48109.}
\thanks{David~J.~States is with the Departments of Bioinformatics and Human Genetics at the University of Michigan, Ann Arbor, MI-48109.}
\thanks{A part of this work has been presented at the Computational Systems Bioinformatics (CSB) 2007 conference, as well as at the Statistical Signal Processing Workshop (SSP) 2007.}
}

\maketitle

\begin{abstract}
Gene regulation is a complex process involving the role of several genomic elements which work in concert to drive spatio-temporal expression.  The experimental characterization of gene regulatory elements is a very complex and resource-intensive process. One of the major goals in computational biology is the \textit{in-silico} annotation of previously uncharacterized elements using results from the subset of known, previously annotated, regulatory elements.

The recent results of the ENCODE project (\emph{http://encode.nih.gov}) presented  in-depth analysis of such functional (regulatory) non-coding elements for $1\%$ of the human genome. It is hoped that the results obtained on this subset can be scaled to the rest of the genome. This is an extremely important effort which will enable faster dissection of other functional elements in key biological processes such as disease progression and organ development (\cite{Kleinjan2005},\cite{Lieb2006}. The computational annotation of these hitherto uncharacterized regions would require an identification of features that have good predictive value.

Gene regulation in higher eukaryotes involves a complex interplay between the gene proximal promoter and distal elements (such as enhancers). Though the exact mechanism of gene regulation is not completely known, several data-driven models have been hypothesized to understand transcription, pointing to sequence, expression, transcription factor (TF) and their interactome level attributes, at  the biochemical level. This has largely been possible due to the advent of new techniques in functional genomics, such as TF chromatin immunoprecipitation (ChIP), RNA interference, microarray yeast-2-hybrid (Y2H) screens. However, these features are yet to be meaningfully integrated for understanding transcriptional regulatory mechanisms. It is believed that such data-driven computational models can be extremely useful to the discovery of new regulatory elements of desired function.

In this work, we study transcriptional regulation as a problem in heterogeneous data integration, across sequence, expression and interactome level attributes. Using the example of the \textit{Gata2} gene and  its recently discovered urogenital enhancers \cite{Khandekar2004} as a case study, we examine the predictive value of  various  high throughput functional genomic assays (from projects like ENCODE and SymAtlas) in characterizing these enhancers and their regulatory role. Observing results from the application of modern statistical learning methodologies for each of these data modalities, we propose a set of features that are most discriminatory to find these enhancers.

\end{abstract}

\begin{keywords}
Nephrogenesis, Random Forests, Transcriptional regulation, Transcription factor binding sites (TFBS), \textit{GATA} genes, comparative genomics, functional genomics, tissue-specific genes, heterogeneous data integration.
\end{keywords}

\section{Introduction}\label{introduction}
\begin{figure}[!h!t!b]
\centerline{\includegraphics[width=3.5in,height=1.5in]{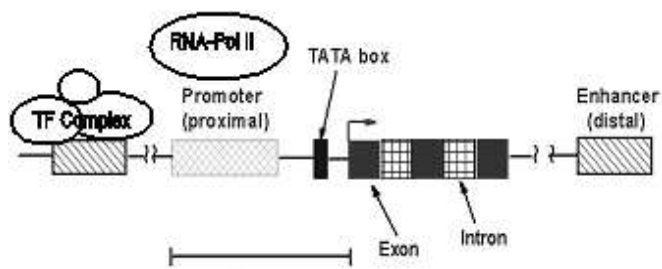}}
\caption{Schematic of Transcriptional Regulation. Sequence motifs at the promoter and the distal regulatory elements together confer specificity of gene expression via TF binding.}\label{fig:transcription}
\end{figure}

Understanding the mechanisms underlying regulation of tissue-specific gene expression remains a challenging question. While all mature cells in the body have a complete copy of the human
genome, each cell type only expresses those genes it needs to carry out its assigned task. This includes genes required for basic cellular maintenance (often called "house-keeping genes") and those genes whose function is specific to the particular tissue type the cell belongs to. Gene expression by way of transcription is the process of generation of messenger RNA (mRNA) from the DNA template representing the gene. It is the intermediate step before the generation of functional protein from messenger RNA. During gene expression, transcription factor (TF) proteins are recruited at the proximal promoter of the gene as well as at sequence elements
(enhancers/silencers) which can lie several hundreds of kilobases from the gene's transcriptional start site (Fig. \ref{fig:transcription}).

It is hypothesized that the collective set of transcription factors that drive (regulate) expression of a target gene are cell, context and tissue dependent (\cite{EnhancerBrowser},\cite{EnhancerBrowser2}). Some of these TFs are recruited at proximal regions such as the promoter of the gene, while others are recruited at more distal regions, such as enhancers. There are several (hypothesized) mechanisms for promoter-enhancer interaction via TF-complex recruitment \cite{looping_scan_track}, by which TFs binding at these regulatory elements could interact during formation of the transcriptional-complex. 

To understand the role of various genomic elements in governing gene regulation, functional genomics has played an enabling role in providing heterogeneous data sources and experimental approaches to discern interactions at the transcriptional, post-transcriptional and translational level. Each of these experiments have aimed to resolve different aspects (features) of transcriptional regulation focussing on TF binding, promoter modeling and epigenetic preferences for tissue-specific expression in some genomic regulatory elements (\cite{ENCODE},  \cite{chromatin-ChIP}, \cite{Sanger-Histone}, \cite{Segal2006}).  Additionally, some studies have demonstrated that these data sets along with principled statistical metrics can be used to understand such features computationally, with a view to asking biologically relevant questions (\cite{chromatin-ChIP},\cite{Segal2006}).

There have been several principled yet scattered studies characterizing the role of regulatory elements such as enhancers for certain genes (such as \textit{Mecp2}, \textit{Shh}, \textit{Gata2}, \textit{Gata3}) in various organisms (\cite{Mecp2},\cite{Shh}) . These are indicative of the inherent spatio-temporal context of gene expression and regulation.
However, there is a need for a unified set of principles underlying the behavior of these enhancers. 
Several models for enhancer-promoter interaction have been published, but it is not really clear what makes a specific genomic element function as a gene-specific enhancer in a certain cellular context. We note that there are promoter-independent enhancers too, and their computational study has been far more principled (\cite{EnhancerBrowser},\cite{Enhancer_Prediction}). 
Several questions arise in this setting - are there any specific sequence properties of such elements, do they harbor/recruit TFs that are expressed highly in that tissue or have a regulatory influence on the target gene discernable at the expression level. Is it possible to determine which TFs are actually recruited from the vast sea that exists at any given time in the cell? - such information along with protein-protein interaction between promoter and enhancer, can yield valuable insight into the behavior of such regulatory elements.



The results of the ENCODE project (\emph{http://encode.nih.gov/}, (\cite{ENCODE},\cite{Sanger-Histone}) on $1\%$ of the human genome has established some very interesting results about the nature of transcriptional regulation at the genome scale. Particularly, they report the use of several experimental techniques (Histone ChIP on chip, DNASE1 hypersensitive assays) etc analyzing transcribed regions as well as their regulatory regions genome-wide. There is now a large scale computational effort developing alongside to ``learn" features of such regulatory elements and use these features for predicting other control elements for genes outside the ENCODE regions, thereby accomplishing a genome-wide annotation. Considering that over  $98\%$ of the genome is non-coding, this annotation effort is going to parallel the previous project in gene-annotation at the genome scale in effort and importance. Adding to this complexity is the fact that the same non-coding element can potentially regulate the expression of genes in a spatio-temporal manner, activating different genes at different times in different tissues, and from arbitrarily large distances from the gene. Thus there is a need for the principled ``reverse-engineering" of the architectures of these regulatory elements, using features at the sequence, expression and interactome level.

Understanding the mechanism of transcriptional regulation thus entails several aspects:
\begin{itemize}
\item
Do regulatory regions like promoters and enhancers have any interesting \emph{sequence properties} depending on the tissues that the corresponding genes are expressed in? These properties are examined based on their individual sequences or their epigenetic preferences. A common technique of analysis is the identification of tissue-specific motif-signatures (\cite{Fraenkel2006}, \cite{Kreiman2004}) for such elements.
\item
Which TFs are recruited at these control elements (promoters and enhancers)? More particularly, is there a correspondence between the motifs representative of the signatures identified from item $1$ and the corresponding TF. Furthermore, is there a method for the principled identification of such TF effectors using modalities such as genome-wide expression data.
\item
Transcriptional regulation is a complex interplay of TF recruitment at sequence motifs on DNA, and their interactions across various regulatory elements (promoters, enhancers, silencers etc.). Given the diversity of the various data sources examining each of these modalities, is there a principled methodology for the integration of these diverse data sources to understand the biology of gene expression?
\end{itemize}


As a case study to ask some of these questions, we examine the regulation of \textit{Gata2} regulation in the developing kidney. \textit{Gata2} is a gene belonging to the GATA family of transcription factors (\textit{GATA1-6}), and has the consensus -WGATAR- motif on DNA \cite{Gata2}. It is located on chromosome $6$, and plays an important role in mammalian hematopoiesis, nephrogenesis and CNS development, with important phenotypic consequences. The study of long-range regulatory elements that effect \textit{Gata2} expression has been on for a couple of years now. The most common strategy for identifying possible regulatory elements has hitherto been inter-species conservation studies. Using this approach, all elements flanking the gene that are conserved more than some threshold and are longer than some limit are retained for further experimental characterization. Given the technical complexity of associated transgenic experiments, this turns out to be a fairly inefficient strategy, especially since the number of candidate regulatory elements increases as the region of comparison, flanking the gene, is expanded (to account for distal regulation). Recently, \cite{Khandekar2004} reported the characterization of two enhancer elements , conferring urogenital-specific expression of \textit{Gata2}, between $80-120$kb away from the gene locus, on chromosome $6$. In this work, we examine, if additional features, at the sequence, expression or interactome level are predictive of the location of these elements, apart from simple sequence comparison. We will also attempt to motivate the utility of these approaches (metrics and data sources) as well as their biological relevance alongside (how they fit into the biophysics of transcriptional regulation). It must be pointed out that there is large paucity in data availability, in that data specific to the developing kidney is hard to come by. Under this constraint, we have made some biologically plausible assumptions so as to obtain maximum information from currently available data sources.


\section{Data Sources:} \label{data_sources}

To understand enhancer regulation of \textit{Gata2} in kidney (based on sequence, expression and interactome perspectives), we utilize data from several data sources:
\begin{enumerate}
\item
To build motif signatures underlying kidney-specific enhancer activity, it would be best to have a database of previously characterized urogenital (UG) enhancers. However, due to the unavailability of such data, we utilize kidney-specific promoter sequences and histone-sequences of enhancers to find motif-signatures of regulatory elements that are potentially UG enhancers.
\begin{itemize}
\item
\underline{Promoters of kidney-specific genes}:
A catalog of kidney-specific mouse promoters is available from the GNF Symatlas (\emph{http://symatlas.gnf.org/}). This database contains list of annotated genes and their expression in several tissue types, including the kidney. Since the proximal promoter of such kidney-specific gens harbors the transcriptional machinery for gene regulation, their sequences putatively have motifs that are associated with kidney-specific expression. Additionally, promoters that are spatio-temporally expressed during kidney development can also be screened (\emph{http://www.informatics.jax.org/}).

We set up the motif discovery as a feature extraction problem from these tissue-specific promoter sequences and then build  a random forest (RF) classifier to classify new promoters into specific and non-specific categories based on the identified sequence features (motifs). Using the RF classifier algorithm we are able to accurately classify more than $98\%$ of tissue specific genes based upon their upstream promoter region sequences alone.
\item
\underline{Chromatin marks in known regulatory elements}: Using the recently released ENCODE data, a catalog of sequences that undergo histone modifications such as methylation and acetylation is available for analysis. Reports suggest that mono-methylation of the lysine residue of Histone $H3$ is associated with  enhancer activity \cite{chromatin-ChIP} whereas tri-methylation of $H3K4$ and $H3$ acetylation are associated with promoter activity. Together, these chromatin marks are indicative of the epigenetic basis of gene transcription. Using the set of $H3K4me1$, $H3K4me3$ and $H3ac$ sequences on chromosome $6$ as training data, we aim to find motifs that are indicative of such epigenetic preferences based only on sequence. We only consider chr:$6$ in order to reduce sequence bias across other chromosomes. Though data is available for five different cell lines, we choose the HeLa cell line data because of its widespread use as a model system to understand transcriptional regulation \textit{in-vitro} in the laboratory. 
With the increasing availability of DNAse1 HS sites \cite{Crawford-HS} for different cell types (\emph{http://research.nhgri.nih.gov/DNaseHS/}) and nucleosome occupancy \cite{Segal2006} data, such analysis is potentially useful for the identification of regulatory controls such as enhancers and promoters. This dataset is referred to as ``histone-modified sequences" in this paper. We note that this data is \emph{not} kidney-specific, since such data is yet to become available. However, the goal in using this data is to find epigenetic preferences from sequence, the idea being to obtain a combined prediction of regulatory elements from data sources $1$ and $2$.
\end{itemize}
\item
Expression data for the developing mouse kidney: there is limited expression data for the developing mouse kidney, mainly due to technical reasons concerning small tissue yield at such early time points. For this study, we use microarray expression data from a public repository of kidney microarray data (\emph{http://genet.chmcc.org}, \emph{http://spring.imb.uq.edu.au/}). Each of these sources contain expression data profiling kidney development from about day $10.5$ dpc to the neonate stage. Some of these studies also examine expression in the developing ureteric bud (UB), metanephric mesenchyme (MM) apart from the whole kidney. This expression data is mined for potential influence between TF genes and \textit{Gata2}, suggesting regulation (Secs:\ref{DTI_theory} and \ref{sec_boot}).
\item
A database of known protein-protein interactions(PPI) : the STRING database (\emph{http://string.embl.de}) integrates various experimental modalities (genomic context, high-throughput experiments such as co-immunoprecipitation, co-expression and literature) to maintain a current list of organism-specific functional protein-association networks. This enables us to explore the interactome-level characteristics of distal enhancer-promoter interaction (Sec:\ref{ppi}).
\end{enumerate}

\section{Rationale}

\begin{figure}[!h!t!b]
\centerline{\includegraphics[width=3.65in,height=4.4in]{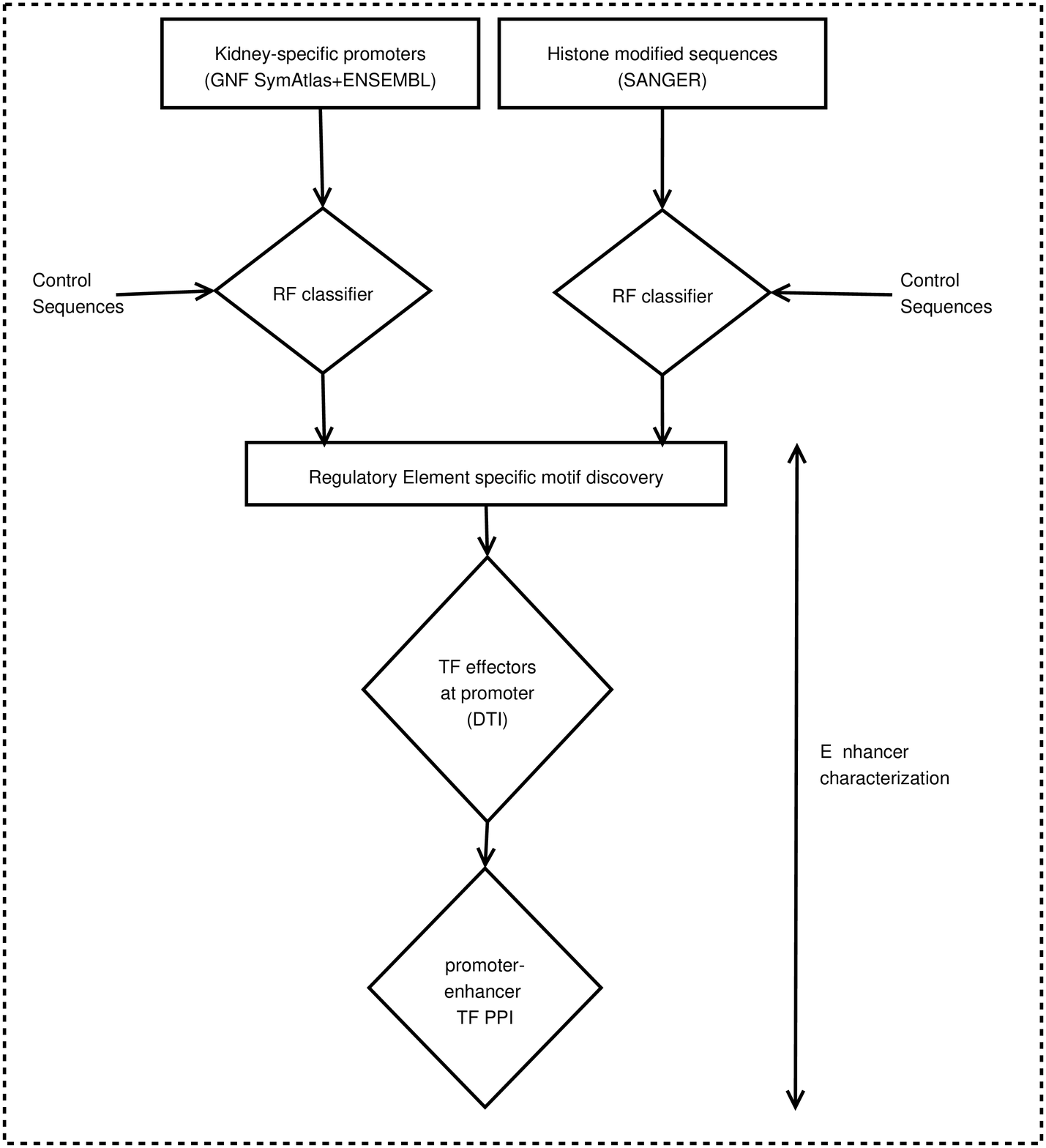}}
\caption{Overall schematic of the proposed methodology.}\label{fig:rationale}
\end{figure}

For the purpose of defining enhancer activity in the developing urogenital system from the various data sources (Sec: \ref{data_sources}), our approach is outlined in Fig. \ref{fig:rationale},

\begin{itemize}
\item
\emph{Feature selection}: In a machine learning context, the identification of sequence-motifs that can discriminate between tissue-specific and non-specific elements (promoters or variably methylated histone  sequences), is a feature selection problem. Here, the features are the counts of sequence-motifs in these training sequences. Without loss of generality, we use six-nucleotide motifs (hexamers) as the motifs. This is based on the observation that most transcription factor binding motifs have a $5-6$ nucleotide core sequence with degeneracy at the ends of the motif.

A similar setup has been introduced in (\cite{DrosophilaFeatureDiff}, \cite{PromFind}). We find that $4^6$ possibilities (from hexamer sequences) yields good performance without being unduly computationally complex. The presented approach, however, does not depend on motif length and can be scaled depending on biological knowledge.
Here, to understand the sequence properties of kidney-specific regulatory elements, we use random forest (RF) classifiers to obtain a sequence of discriminating hexamer motifs between kidney-specific promoters and housekeeping promoters. Additionally, we build a RF classifier to discriminate monomethlyated $H3K4$ sequences from trimethylated $H3K4$/acetylated $H3$ sequences. This yields motifs associated with epigenetic properties of promoters and enhancers, which are potentially predictive of regulatory potential for novel sequences.


\item
\emph{TF Influence determination}: After discovering key discriminating motifs using the above RF step, we examine the discovered motifs for matches with known transcription factor binding site profiles at the \textit{Gata2} promoter. For those that match known TFBS, we look for computational evidence of a directed influence from the TF encoding gene to the gene of interest (here, \textit{Gata2}) based on microarray expression data \cite{CSB2007}. This seeks to integrate sequence and expression data into the determination of transcription factor-target relationships.

\item
\emph{Examining promoter-enhancer TF interactions using PPI interactome:} The identification of phylogenetically conserved effector TFs at the promoter (identified via DTI) can lead to the exploration of interactions between these TFs and those that are phylogenetically conserved at the UG enhancers, borrowing from their expected interaction modes during long range regulation \cite{looping_scan_track}.
\end{itemize}

In this work, all these questions will be integratively answered for training data as well as in the context of the urogenital enhancers identified in \cite{Khandekar2004}. We aim to show that each of these `features'  has a predictive value for the identification of enhancers and the integration of these heterogeneous data can lead to potential reduction in false positive rate during large-scale enhancer discovery, genome-wide. To date, there has been no comprehensive study for summarizing various heterogeneous data sources to understand transcriptional regulation.





The main approaches to finding motifs relevant to certain classes with respect to examining common motifs driving gene regulation are
summarized in (\cite{Kreiman2004}, \cite{Fraenkel2006}). The most common approach is to look for TFBS motifs (TRANSFAC / JASPAR) that
are statistically over-represented based. This assumes a parametric form (Binomial/Poisson) on the probability density of motifs in the
population of promoters of co-expressed genes.

We set-up the problem of discriminative motif discovery as a word-document classification problem. Having constructed two groups
of genes for analysis, tissue specific (\textit{'ts'}) and non-tissue specific (\textit{'nts'}) - we seek to find hexamer
motifs which are most discriminatory between these two classes. Our goal would be to make this set of motifs as small as possible - i.e.
to achieve maximal class partitioning with the smallest feature subset. Towards this goal, we explore the use of random forests (RF)
for finding such a discriminative hexamer subset.

As can be expected, the input to such an approach would be a gene promoter - motif frequency table (Table~\ref{motif_promoter_mtx}).
The genes relevant to each class are identified from tissue microarray analysis, and the frequency table is built by parsing the
gene promoters for the presence of each of the $4^6 = 4096$ possible hexamers.

%
%

\section{Validation/Biological Application}
%
%
%

As suggested in Sec: \ref{introduction}, we use the recently identified \textit{Gata2} urogenital (UG) enhancers to validate our approaches. All the data sources (and its analysis) is therefore going to be centered around the kidney.

To find these elements experimentally, the following strategy was adopted. Based on BAC transgenic \cite{Khandekar2004} studies, the approximate location of the urogenital enhancer(s) of \textit{Gata2} were localized to a $200$ kilobase region on chromosome $6$. Using inter-species conservation plots, four elements were selected for transgenic analysis in the mouse. These were designated UG$1$,$2$,$3$ and $4$. After a lengthy and resource-intensive experimental effort, the UG enhancers were found to be two out of these four non-coding elements, $UG2$ and $UG4$. Our problem takes motivation from this setting - we ask if presently available functional genomic data at the sequence, expression and interactome level could enable the principled discovery of these elements, computationally?

It is easy to see the utility of such a methodology, because such methods can be scaled up contextually for other genes of interest. Given the complexity of $1\%$ of the genome, made possible by the ENCODE project, the search for functional elements genome-wide is going to be an important and challenging exercise. Thus our goal is to find predictive ``features" at the sequence, expression and interactome level based on available data sources and ask if they predict that $UG2$ and $UG4$ are indeed functional in the kidney, whereas $UG1$ and $UG3$ are not.

\section{Organization}

With a view to understanding the elements of transcriptional regulation, the first part of this paper (Sections \ref{data_processing}-\ref{histone-RF}) addresses the problem of identifying motif signatures representative of transcriptional control from kidney-specific promoters and epigenetically marked sequences.  The second part of this work (Sections \ref{DTI_theory}-\ref{sec_boot}) integrates phylogeny and expression data to find regulatory TFs at the proximal promoter of \textit{Gata2}. Using these two pieces, we examine if sequence, expression and protein-interaction data (Sec: \ref{ppi}) can offer supporting evidence for the observed \textit{in-vivo} behavior of four putative \textit{Gata2} regulatory elements. At each step, a validation of the obtained features with $UG1-4$ will be done.

\section{Sequence Data Extraction and Pre-processing}\label{data_processing}

The Novartis foundation tissue-specificity atlas [\emph{http://symatlas.gnf.org/}], has a compendium of genes and their corresponding tissues of expression. Genes have been profiled for expression in about twenty-five tissues, including adrenal gland, brain, dorsal root ganglion, spinal chord, testis, pancreas, liver etc. If a gene is expressed in less than three tissue types, it is annotated tissue-specific (\textit{`ts'}), and if it is expressed in more than $22$ tissue types, it is annotated to be non-specific (\textit{`nts'}). Based on this assignment, we find a list of $86$ genes that are tissue-specific as well as have kidney expression (MGI:\emph{http://www.informatics.jax.org/}). For these kidney-specific genes, we extract their promoter sequences from the ENSEMBL database \emph{http://www.ensembl.org/}], using  sequence $2000$bp upstream and $1000$bp downstream up to the first exon relative to the transcriptional start site reported in ENSEMBL (release 37).

Before proceeding to motif selection, a matrix of motif-promoter correspondences is created. In this matrix, the counts of hexamer (six-nucleotide) motif occurrence in the \textit{`ts'} and
\textit{`nts'} promoters is obtained using sequence parsing. The motif length of six is not overly restrictive, since it corresponds to the consensus binding site size of several
annotated transcription factor motifs in the TRANSFAC/JASPAR databases. A Welch t-test is then performed between the relative counts of each hexamer in the two expression categories (\textit{`ts'} and \textit{`nts'}) and the top $1000$ hexamers with $p-value \le 10^{-6}$ are selected. This set of discriminating hexamers is designated ($\overrightarrow{\textbf{H}} =
H_1,H_2,\ldots,H_{1000}$). This procedure resulted in two hexamer-gene co-occurrence matrices, - one for the \textit{`ts'} (or $+1$) class of dimension $N_{train,+1} \times 1000$ and the
other for the \textit{`nts'} (or $-1$) class - dimension $N_{train,-1} \times 1000$. Here $N_{train,+1}$ is the matrix of the $86$ kidney-specific genes. $N_{train,-1}$ is the set of \textit{`nts'} that do not have kidney-specific expression.

As an illustration, we show a representative matrix (Table $1$).
\begin{table}[!h!b!t]
\centering
\begin{tabular}{cccccc}
Ensembl Gene ID & AAAAAA    & AAATAG  & Class \\
ENSG00000155366 & 1      & 1   &  +1 \\
ENSG000001780892 & 4   & 3 &   +1 \\
ENSG00000189171 & 1  &  2 &    -1 \\
ENSG00000168664 & 4  &  3 &   -1 \\
ENSG00000160917 & 2  &  1 &    -1 \\
ENSG00000176749 & 1  &  1 &   -1 \\
ENSG00000006451 & 3  &  2 &   +1
\end{tabular}
\caption{The 'motif frequency matrix' for a set of gene-promoters.
The first column is their ENSEMBL gene identifiers, the next $2$
columns are hexamer quantile labels, and the last column is the
corresponding gene's class label ($+1/-1$).}\label{motif_promoter_mtx}
\end{table}

All the above steps, from promoter sequence extraction, parsing and quantization to obtain hexamer-promoter counts that are done for the kidney-specific genes can be repeated for the histone-modified sequences. This dataset is obtained from the Sanger ENCODE database (\emph{http://www.sanger.ac.uk/PostGenomics/encode/data-access.shtml}), and contains $681$ sequences that undergo modification ($m1/me3/ac$) in histone ChIP assays. $337$ of these correspond to $H3K4me1$ (enhancers), and $344$ correspond to $H3K4me3/H3ac$ marks (promoters). Here, the $1000$ hexamers discriminating $H3K4me1$-sequences ($+1$ set) and a $(H3K4me3/H3ac)$ ($-1$), are designated $\overrightarrow{\textbf{H'}}=H'_1,H'_2,\ldots,H'_{1000}$.


\begin{table}[!h!b!t]
\centering
\begin{tabular}{cccccc}
Sequence & AAAATA & AAACTG & Class\\
chr6:41410492-41411867 & 2 & 1 & +1 \\
chr6:41654502-41654782 & 4 & 2 & +1 \\
chr6:41406971-41408059 & 1 & 1 & -1 \\
chr6:41665970-41667002 & 2 & 3 & +1 \\
chr6:41476956-41478365 & 1 & 2 & -1 \\
chr6:41530471-41531046 & 2 & 2 & -1 \\
chr6:41783327-41784532 & 1 & 2 & +1
\end{tabular}
\caption{The 'motif frequency matrix' for a set of histone-modified sequences.
The first column is their genomic locations along chr$6$, the next $2$
columns are hexamer quantile labels, and the last column is the
corresponding sequence class label ($+1/-1$).}\label{motif_histone_mtx}
\end{table}

\section{Motif-Class Correspondence Matrices}

From the above, $N_{train,+1} \times 1000$ and $N_{train,-1} \times 1000$ dimensional co-occurrence matrices are available for the tissue-specific and non-specific data, both for the promoter and histone-modified sequences. Before proceeding to the feature (hexamer motif) selection step, the counts of the $M = 1000$ hexamers in each training sample need to be normalized to account for variable sequence lengths. 
In the co-occurrence matrix, let $gc_{i,k}$ represent the absolute count of the $k^{th}$ hexamer, $k \in {1,2,\ldots,M}$ in the $i^{th}$ gene. Then, for each gene $g_{i}$, the quantile labeled matrix has $X_{i,k} = l$ if $gc_{i,[\frac{l-1}{K}M]} \le gc_{i,k} < gc_{i,[\frac{l}{K}M]}, K=4$. Matrices of dimension $N_{train,+1} \times 1001$, $N_{train,-1} \times 1001$ for the specific and non-specific training samples are now obtained. Each matrix
contains the quantile label assignments for the $1000$ hexamers $(X_i, i \in (1,2,\ldots,1000))$, as stated above, and the last column would have the corresponding class label ($Y = -1/+1$). 

\section{Random Forest Classifiers}

A random forest (RF) is an ensemble of tree classifiers obtained by aggregating (bagging) several classifiers, mostly classification trees. Such classifiers have provably low bias and variance characteristics and are extremely amenable to random data subset selection via bootstrapping. In a RF approach, an ensemble of classification trees is built on a training set and validated on an out of bag (OOB) testing set. As compared to ordinary decision tree
classifiers where only one variable is used to split the node optimally, random forests allow the use of a variable subset that optimally split each node leading to a much cleaner class discrimination at every node. The variables selected for optimal partitioning over class labels can be examined from a variable importance plot which indicates which variables are most discriminatory between these two classes \cite{Breiman_RF}. It is also to be noted
that unlike most classifiers, which require a separate cross-validation procedure, random forests afford the dual advantage of training and cross-validation (through the OOB data) during the training procedure. Thus each tree is multiply cross-validated before being incorporated into the classifier ensemble.


Several interesting insights into the data are available using random forests. The variable importance plot yields the variables that are most discriminatory for classification under the `ensemble of trees' classifier. This importance is based on two measures- `Gini index' and `decrease in accuracy'. The Gini index is an entropy based criterion which measures the purity of a node in the tree, while the other metric simply looks at the relative contribution of each variable to the accuracy of the classifier. The performance of the classifier is visualized with receiver operating characteristic (ROC) curves, by plotting the true positive rate against the false positive rate. The best classifier has the co-ordinates $(0,1)$ on the ROC plot. For our studies, we use the `randomForest' package for R \cite{Breiman_RF}. The classifier performance on the individual data and the related diagnostics are mentioned under each head (Secs: \ref{kidney-RF} and \ref{histone-RF}).

\section{Random Forests on Kidney-specific promoters}\label{kidney-RF}
In this section, we aim to find discriminating sequence motifs between a set of kidney-specific promoters and housekeeping promoters with a goal to find sequence motifs underlying kidney-specific regulation. The kidney enriched dataset has $86$ genes that are assigned to a tissue specific class and have higher than mean expression in the kidney. For the purpose of training and testing, we consider another set of genes that are not tissue-specific in the kidney. Using this approach, we obtain a classification accuracy of $> 95\%$ on the kidney enriched tissue specificity data set.

\begin{figure}[!h!b!t]
\centerline{\includegraphics[width=3.2in,height=3in]{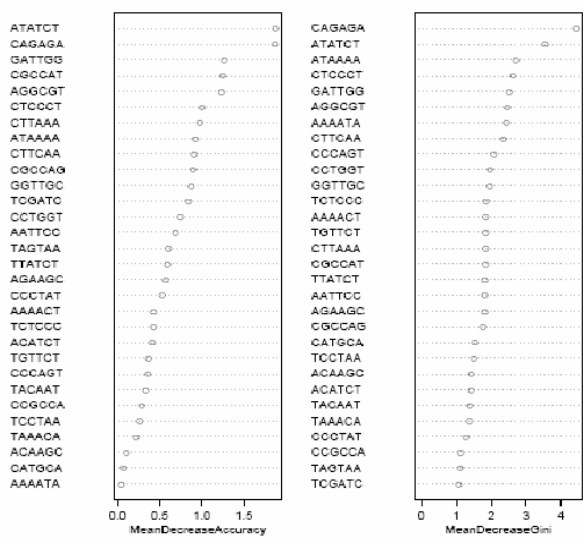}}
\caption{Top hexamers which can discriminate between kidney-specific
and house-keeping genes.}\label{fig:kidney_hexamers}
\end{figure}

Before proceeding to motif identification, it is necessary to check for possible sequence bias (GC composition) between the two classes of promoters (kidney-specific vs. housekeeping). If there is a significant bias, then the motifs turn out to be just GC rich sequences that are not very biologically informative \cite{DWE2005} for regulatory potential. The GC composition of these two classes of sequences is represented in Fig. \ref{fig:kidney_gc}.  As can be seen, the average GC composition is the same. The ROC and variable importance plot for the overall classification is indicated below (Fig. \ref{fig:roc_rf} and \ref{fig:kidney_hexamers}).



%


To address a related question, we examine if the top ranked hexamers in the kidney dataset correlate sequence-wise with known transcription factor binding sites. Using the publicly available Opossum tool (\textit{http://www.cisreg.ca/cgi-bin/oPOSSUM/opossum/}) or MAPPER (\emph{http://bio.chip.org/mapper}), we found several interesting transcription factors to map to these motifs, such as \textit{Nkx}, \textit{ARNT}, \textit{c-ETS}, \textit{FREAC4}, \textit{NFAT}, \textit{CREBP}, \textit{E2F}, \textit{HNF4A}, \textit{Pax2}, \textit{MSX1}, \textit{SP1} several of which are kidney-specific. Though this is highly consistent with the dataset, the functional relevance of  these sites remains to be experimentally validated.

%

\section{RFs on chromatin-modified sequences}\label{histone-RF}

We train a RF classifier on a set of $681$ sequences from chromosome $6$ that have varying histone modifications associated with them (namely, $H3K4me1/me3$, and $H3ac$ ), as mentioned in Section: \ref{data_sources}. These are derived from the HeLa cell line and are not necessarily context-specific for kidney development. However, given the widespread  use of this cell line for transcriptional studies, we aim to find if the motifs associated with regulatory elements are indeed predictive of enhancer activity.

\begin{figure}[!t!h!b]
\centerline{\includegraphics[width=3.2in,height=3in]{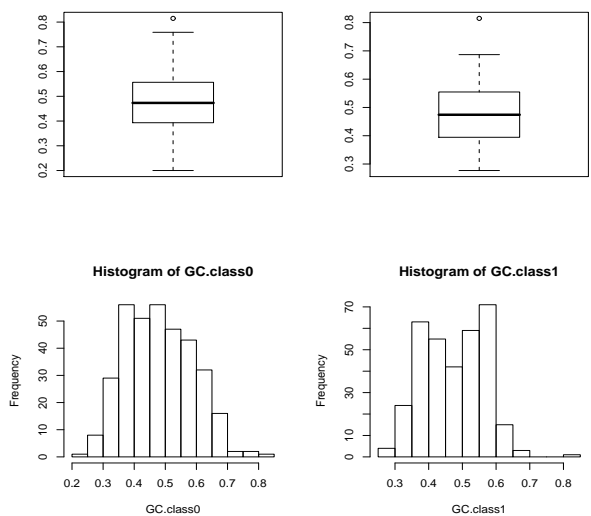}}
\caption{GC plots for sequence bias in $H3K4me1$ histone sequences vs. $H3K4me3$ and $H3ac$  sequences.}\label{fig:histone_gc}
\end{figure}

Here too, we examine the GC-composition bias of these two sequence classes (Fig. \ref{fig:histone_gc}) and confirm that there is no such sequence bias that would skew the discovery and subsequent interpretation of these epigenetic motifs.

\begin{figure}[!t!h!b]
\centerline{\includegraphics[width=3.2in,height=3in]{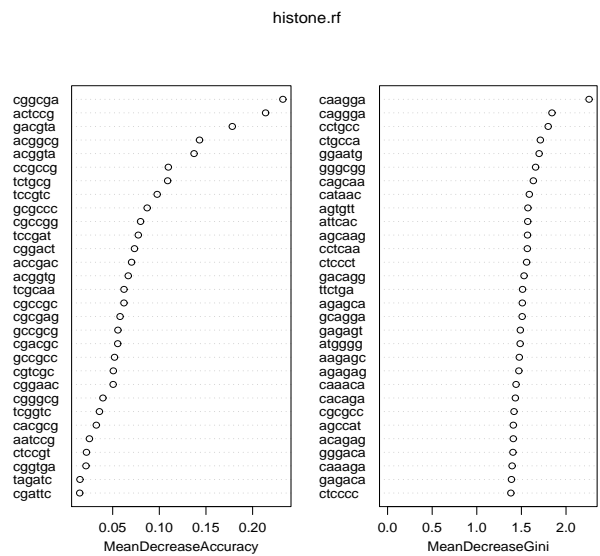}}
\caption{Top hexamers which can discriminate between $H3K4me1$ histone sequences vs. $H3K4me3$ and $H3ac$ sequences.}\label{fig:histone_hexamers}
\end{figure}

\begin{figure}[!t!h!b]
\centerline{\includegraphics[width=3in,height=1.6in]{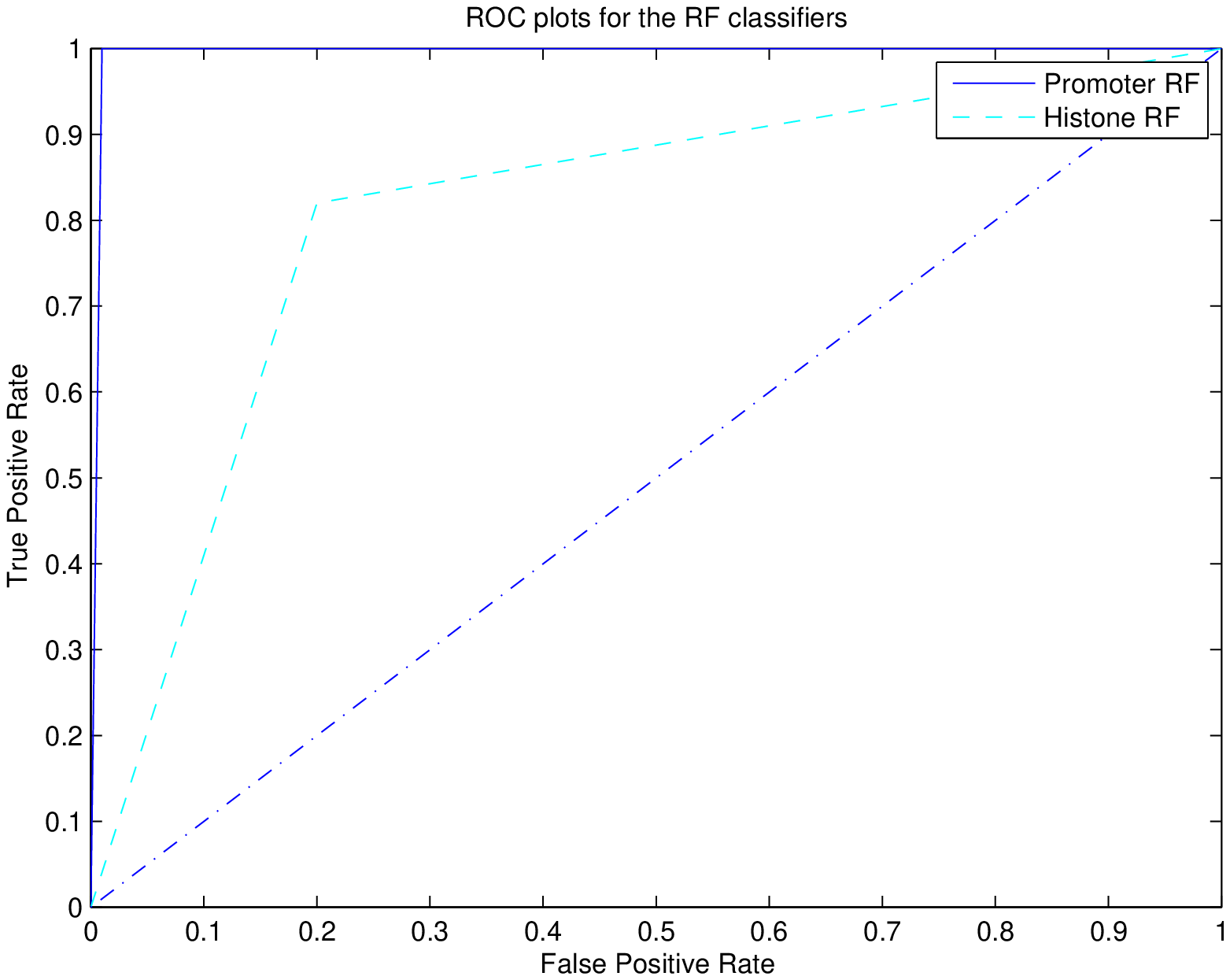}}
\caption{ROC plots for the two RF classifiers (RF-promoter in solid, and RF-histone in dashed line). The diagonal line is the classification by random chance.}\label{fig:roc_rf}
\end{figure}

The ROC plots for the two random forest classifiers is given in Fig. \ref{fig:roc_rf}. As can be seen, the kidney-promoter based classifier has a much superior performance than the histone modification-based classifier. However these are two complementary data sources and can be effectively combined to improve detection reliability.

The motifs obtained from the random forest analysis indicate the sequence preferences of regulatory elements that are kidney-specific or nucleosome-free. We analyze the performance of these classifiers on the $4$ UG enhancers, mentioned previously. In both cases $UG2-4$ are classified as kidney-specific enhancers, whereas $UG1$ is correctly classified as not being regulatory. Additionally, a control set of enhancers derived from the Mouse Enhancer database was also classified as enhancers based on these chromatin signatures. This high prediction accuracy inspite of non-specificity of cell context 
is very interesting and has potentially high predictive value. However, the higher false positive rate (indicated in the ROC plot) can be explained based on the fact that these sequences were derived from a cell population  that was not kidney-specific.  


\section{DTI formulation}\label{DTI_theory}

Since our goal is to understand the nature of long-range transcriptional regulation, we can examine the role of these discovered motifs 
using expression and interactome data. The first question that arises in this context is if any of these discovered sequence motifs (from kidney-specific or histone modification sequences) are related to \textit{Gata2} transcription at the expression level. Additionally, this can help resolve which TFs bind at these regulatory elements as well as if there is an interaction between them that underlies tissue specific regulation/gene expression. Recently, we introduced the directed information (DTI) as a metric to infer expression-level influence between any putative transcription factor (TF) gene and a target gene (such as \textit{Gata2}) \cite{CSB2007}. We will briefly summarize the utility of DTI for TF effector identification in these sections (Sec. \ref{DTI_theory} and \ref{sec_boot}).


%

Using inter-species conservation and TFBS matching databases (TRANSFAC/JASPAR) we can find the transcription factors that putatively bind to the \textit{Gata2} promoter. Using publicly available expression data for the developing kidney (\cite{GrimmondLittle2005}, \cite{Stuart2003}), we can find TF effectors from this conserved set as well as from TFs corresponding to top ranking classifier motifs. 


The DTI is a directed dependence metric that quantifies the influence between a putative TF effector ($X$) and \textit{Gata2} ($Y$), based on mRNA expression data. Briefly, the DTI (for a lag of 1) between two $N$-length random processes $X$ and $Y$ is given by \cite{Massey1990} :
\begin{align*}
I(X^N \rightarrow Y^N) = \sum_{n=1}^{N}I(X^n;Y_n|Y^{n-1}) \tag{$1$}
\end{align*}
Here, $Y^n$ denotes $(Y_1, Y_2,.., Y_n)$, i.e. a segment of the realization of a random sequence $Y^n$ and $I(X^n;Y^n)$ is the Shannon mutual information . As already known, $I(X^n; Y^n) = H(X^n)-H(X^n|Y^n)$, with $H(X^n)$ and $H(X^n|Y^n)$ being the Shannon entropy of $X^n$ and the conditional entropy of $X^n$ given $Y^n$, respectively. Using this definition of mutual information, the Directed Information simplifies to,

\begin{gather*}
I(X^N \rightarrow Y^N) = \sum_{n=1}^N [H(X^n|Y^{n-1})-H(X^n|Y^{n})] \\
= \sum_{n=1}^N \{ [H(X^n,Y^{n-1})-H(Y^{n-1})]-
[H(X^n,Y^{n})-H(Y^{n})]\} \tag{$2$}
\end{gather*}

To infer the notion of influence between two time series (mRNA expression data) we find the mutual information between the entire evolution of gene $X$ (up to the current instant $n$) and the current instant of $Y$ ($Y_n$), given the evolution of gene $Y$ up to the previous instant $n-1$ (i.e. $Y^{n-1}$). This is done for every instant $n \in (1,2,\ldots,N)$ in the $N$ - length expression time series. Thus, we find the influence relationship between genes
$X$ and $Y$ for every instant during the evolution of their individual time series. 

As can be seen, this computation requires the estimation of joint and marginal entropies, which are done via data-dependent partitioning of the observation space (\cite{Darbellay-Vajda1999}, \cite{ErikLearnedMiller2003}).  Replicate (biological, technical and probe-level) gene expression data is very useful for this purpose and enables entropy estimation from moderate
sample size. Additionally, several methods exist for entropy estimation from moderate sample sizes. One of the most prominent is the Voronoi tessellation approach outlined in \cite{ErikLearnedMiller2003}. In this approach, an adaptive partitioning of the observation space is used to estimate the probability densities as well as the entropies of the random variables.

From the definition of DTI, we know that $0 \le I(X_i^N \rightarrow Y^N) \le I(X_i^N;Y^N) < \infty$ .For easy comparison with other metrics, we use a normalized DTI metric \cite{John1989b} given by, $\rho_{DI}= \sqrt{1-e^{-2 I(X^N \rightarrow Y^N)}} = \sqrt{1-e^{-2\sum_{i=1}^N I(X^i;Y_i|Y^{i-1})}}$. This maps the large range of DI, ($[0,\infty]$) to lie in $[0,1]$. Another point of consideration is to estimate the significance of the DTI value compared to a null distribution on the DTI value (i.e. what is the chance of finding the DTI value by chance from the series $X_i$ and $Y$). This is done using confidence intervals after permutation testing (Sec: \ref{sec_boot}). We use a threshold $p$-value of $0.05$ to estimate the significance of the true DTI value in conjunction with the the density estimation of a random data permutation, as outlined below. These aspects are explained in \cite{CSB2007}, and are only mentioned below for completeness.



\section{Bootstrapped Confidence Intervals}\label{sec_boot}
In the absence of knowledge of the true distribution of the DTI estimate, an approximate confidence interval for the DTI estimate ($\hat{I}(X^N \rightarrow Y^N)$), is found using bootstrapping \cite{EffronTibshirani1994}. Density estimation is based on kernel smoothing over the bootstrapped samples \cite{Silverman1997}.

The kernel density estimate for the bootstrapped DTI (with $n = 1000$ samples), $Z
\triangleq \hat{I}_B(X^N \rightarrow Y^N)$ becomes,\\
$\hat{f}_h(Z) = \frac{1}{nh}\sum_{i=1}^{n}\frac{3}{4} [1-(\frac{z_i-z}{h})^2] I(\left\vert \frac{z_i-z}{h} \right\vert \le 1)$ with $h \approx 2.67\hat{\sigma}_z$ and $n=1000$.
$\hat{I}_B(X^N \rightarrow Y^N)$ is obtained by finding the DTI for each random permutation of the $X$, $Y$ series, and performing this permutation $B$ times. As is the clear from the above expression, the Epanechnikov kernel is used for density estimation from the bootstrapped samples. The choice of the kernel is based on its excellent characteristics - a  compact region of support, the lowest AMISE (asymptotic mean squared error) and favorable bias-variance tradeoff \cite{Silverman1997}.

We denote the cumulative distribution function (over the bootstrap samples) of $\hat{I}(X^N \rightarrow Y^N)$ by $F_{\hat{I_B}(X^N \rightarrow Y^N)}(\hat{I_B}(X^N \rightarrow Y^N))$. 
 Let the mean of the bootstrapped null distribution be $I_B^{*}(X^N \rightarrow Y^N)$. We denote by $t_{1-\alpha}$, the $(1-\alpha)^{th}$ quantile of this distribution i.e.
\{$t_{1-\alpha}: P([\frac{\hat{I_B}(X^N \rightarrow Y^N)-I_B^{*}(X^N \rightarrow Y^N)}{\hat{\sigma}}] \leq t_{1-\alpha}) = 1-\alpha$\}. Since we need the true $\hat{I}(X^N
\rightarrow Y^N)$ to be significant and close to 1, we need $\hat{I}(X^N \rightarrow Y^N) \geq [I_B^{*}(X^N \rightarrow Y^N)+ t_{1-\alpha} \times \hat{\sigma}]$, with $\hat{\sigma}$ being the standard error of the bootstrapped distribution, \\
$\hat{\sigma} = \sqrt{\frac{[\Sigma_{b=1}^B \hat{I_b}(X^N \rightarrow Y^N)-I_B^{*}(X^N \rightarrow Y^N)]^2}{B-1}}$; $B$ is the number of bootstrap samples.

As an example, we indicate the significance and strength of the DTI between the \textit{Pax2} TF and \textit{Gata2}. The high strength of influence and its significance coupled with the phylogenetic conservation of the \textit{Pax2} motif indicates expression evidence for the role of \textit{Pax2} in \textit{Gata2} regulation (\cite{GrimmondLittle2005},\cite{pax2_gata2_pronephros}).
\begin{figure}[!t!h!b]
\centerline{\includegraphics[width=3in,height=1.6in]{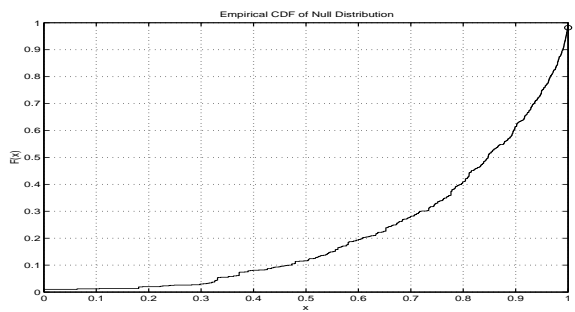}}
\caption{Cumulative Distribution Function for bootstrapped
$I(\textit{Pax2} \rightarrow \textit{Gata2})$ interaction. True $\hat{I}
(\textit{Pax2} \rightarrow \textit{Gata2}) = 0.9818$.
}\label{fig:pax2_gata2}
\end{figure}

\begin{figure}[!t!h!b]
\centerline{\includegraphics[width=3in,height=1.6in]{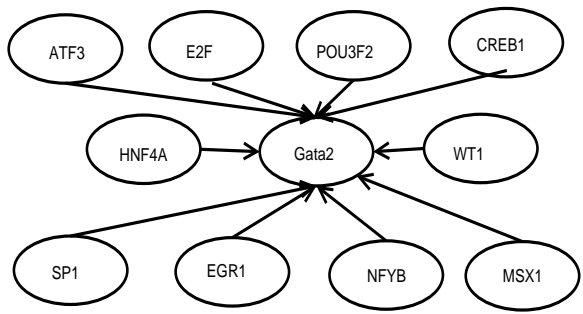}}
\caption{Putative upstream TFs using DTI for the \textit{Gata3}
gene.}\label{fig:effector_gata2}
\end{figure}

Such analysis can be extended to all TFs that are phylogenetically conserved or those that correspond to top-ranking classifier motifs. For \textit{Gata2} UG regulation, one such network is Fig. \ref{fig:effector_gata2},

\section{Protein-Protein Interactions}\label{ppi}


The discovery of putative TF effectors that are involved in \textit{Gata2} expression (identified from a combination of motif signatures and expression DTI) can lead to interesting insights into transcriptional regulatory mechanisms. From \cite{looping_scan_track} previous literature on the nature of long-range transcriptional regulation, we can examine the evidence of interaction between such TFs at the \textit{Gata2} promoter with those at the UG enhancers, and subsequently use such interaction models as predictors of new regulatory elements.

Using a notion of protein-protein interaction to mediate long-distance interactions between promoters and enhancers, we explore the interactome to look for network linkage between the TFs at the promoter (regulatory TFs found from motif search and DTI) and those phylogenetically conserved TFs at the enhancer(s). These interactions are summarized below,

\begin{figure}[!t!h!b]
\centerline{\includegraphics[width=3in,height=1.6in]{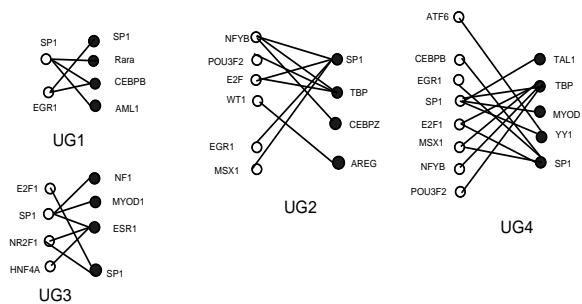}}
\caption{Protein-protein interaction between putative \textit{Gata2} TFs (hollow circles) and putative UG element TFs (filled circles). From \emph{http://string.embl.de/}}\label{fig:pax2_gata2}
\end{figure}

The above figure indicates a very interesting property of the real enhancers vis-a-vis the other conserved elements. We see that the TF effectors for \textit{Gata2} such as \textit{SP1}, \textit{POU3F2} (identified in the TF effector network above, Fig. \ref{fig:effector_gata2}), are involved in cross-element interactions at the protein level, between the promoter and true enhancer ($UG2/4$). However, the network linkage in the elements that showed no enhancer activity is very sparse suggesting low cross-talk between promoter and enhancer. Also, the TFs at the enhancer nodes (dark circles), therefore, have more hubs in the functional elements $UG2/4$ as compared to the non-functional ones. Thus, the extent of cross-talk is a potential discriminator of possible enhancer function.  This shows that superimposing PPI information along with sequence and expression data helps reduce the number of false positives while integrating various aspects of distal regulation. A quantitative metric that summarizes this extent of cross-talk would greatly facilitate in-depth analysis of long-range interaction.

\section{Summary of Algorithm}
%
%
%

Based on the presented data from ENCODE, Enhancer Browser, and SymAtlas sources, we believe that the following features are predictive of regulatory element location:
\begin{itemize}
\item
Motif signatures are predictive of regulatory element location. These comprise signatures derived from tissue-specific gene promoter sequences as well as sequences with various chromatin marks or modifications.  
\item
TFs that are putatively active in gene (\textit{Gata2}) regulation can be identified using a combination of expression data, and tissue-specificity data. 
\item
Effector TFs (via DTI) at the gene proximal promoter have high network linkage with enhancer TFs in case of functional enhancers. Several enhancer TFs are hubs that mediate formation of the transcription factor complex.
\end{itemize}

It is to be noted that this model is data driven and may not directly correspond to the biology of transcription. However, much like markov models for gene sequence annotation, we believe that such data-driven models are useful for genome-wide study.


\section{Conclusions}

In this work, we have examined the problem of regulatory element identification. Such an effort has implications to understand the genomic basis of key biological processes such as  development and disease. Using the biophysics of transcription, this can be modeled as a problem in data integration over various experimental modalities such as sequence, expression, transcription factor binding and interactome-data. Using the case study of enhancers corresponding to the \textit{Gata2} gene, we examine the utility of these heterogeneous data sources for predictive feature selection, using principled methodologies and metrics.

Based on motif signatures, we find that they predict the true enhancers ($UG2$, $UG4$), and the false enhancer $UG1$, but mispredict $UG3$ to be an enhancer. However, superimposing TF effector discovery and protein-protein interaction data yields some heuristics for enhancer discovery based on long range interaction between promoter and enhancer, thereby improving on prediction accuracy.

%
%
%
%
%
%
%
%

\section{Future Work}

Some key elements directly emerge for guiding future research. As already alluded to in the motif-signature procedure, specific expression data corresponding to stages and tissues of interest would greatly improve the specificity of regulatory element prediction.  Furthermore, as histone modification maps for different cell lines  are generated, the false positive rate of prediction would decrease, thereby improving accuracy. Several other learning paradigms can be introduced into this setting, since we are learning from structured data. Conditional random fields have proved to invaluable in such analysis. Also, methods in joint classifier and feature optimization might likely improve the accuracy of predictions.

At the expression level, methods for supervised network inference would have a great impact on the discovery of TF effectors. Rapid advances have been made in this area and their relevance to the biological context of the problem has become very principled. At the interactome level, a metric to quantify the degree of ``connectedness" of the TFs between the enhancer and promoter would be very useful for the construction of a ``interactome-classifier". Other methods that can account for different types of long-range interactions would be extremely useful too.

%
%
%
%
%

\section{Acknowledgements}
The authors gratefully acknowledge the support of the NIH under award 5R01-GM028896-21 (J.D.E). We would also like to thank Prof. Sandeep Pradhan and Mr. Ramji Venkataramanan for useful discussions on directed information. We are extremely grateful to Prof. Erik
Learned-Miller and Dr. Damian Fermin for sharing their code for
high-dimensional entropy estimation and ENSEMBL sequence
extraction, respectively. 

\section{Availability}
The source code of the analysis tools (in R $2.0$ and MATLAB $6.1$)
is available on request. Supplementary data (the set of various tissue-specific promoters, enhancers, housekeeping promoters and hexamer lists for each category) will be made available from the publisher's website.



{\footnotesize
\begin{thebibliography}{}
\addtolength{\baselineskip}{-.1\baselineskip}

\bibitem{Toucan2005}
Aerts S, Van Loo P, Thijs G, Mayer H, de Martin R, Moreau Y, De Moor
B., "TOUCAN 2: the all-inclusive open source workbench for
regulatory sequence analysis", {\em Nucleic Acids Res}. 2005 Jul
1;33(Web Server issue):W393-6.
\bibitem{Breiman_RF}
L. Breiman., "Random forests"., {\em Machine Learning}, 45(1):
5.32, 2001.
\bibitem{Burge1997}
Burge C, Karlin S, "Prediction of complete gene structures in human
genomic DNA". {\em J Mol Biol} 1997, 268:78-94.
\bibitem{GrimmondLittle2005}
Challen G, Gardiner B, Caruana G, Kostoulias X, Martinez G, Crowe M,
Taylor DF, Bertram J, Little M, Grimmond SM.,"Temporal and spatial
transcriptional programs in murine kidney development".,{\em Physiol
Genomics}. 2005 Oct 17;23(2):159-71.
\bibitem{DrosophilaFeatureDiff}
Chan BY, Kibler D., "Using hexamers to predict cis-regulatory motifs
in Drosophila", {\em BMC Bioinformatics}, 2005 Oct 27;6:262.
\bibitem{Crawford-HS}
Crawford GE, Holt IE, Whittle J, Webb BD, Tai D, Davis S, Margulies EH, Chen Y,
Bernat JA, Ginsburg D, Zhou D, Luo S, Vasicek TJ, Daly MJ, Wolfsberg TG, Collins
FS.,"Genome-wide mapping of DNase hypersensitive sites using massively parallel
signature sequencing (MPSS)",.{\em Genome Res}. 2006 Jan;16(1):123-31.
\bibitem{Dressler1992}
Dressler, G.R. and Douglas, E.C. (1992)., "Pax-2 is a DNA-binding
protein expressed in embryonic kidney and Wilms tumor"., {\em Proc.
Natl. Acad. Sci. USA} 89: 1179-1183.
\bibitem{pax2_gata2_pronephros}
Drummond IA.,"The zebrafish pronephros: a genetic system for studies of kidney development"., {\em Pediatr Nephrol}. 2000 May;14(5):428-35.
\bibitem{EffronTibshirani1994}
Effron B, Tibshirani R.J, An Introduction to the Bootstrap
(Monographs on Statistics and Applied Probability), Chapman \&
Hall/CRC, 1994.
\bibitem{ENCODE}
ENCODE Project Consortium, "Identification and analysis of functional elements in 1\% of the human genome by the ENCODE pilot project"., {\em Nature}. 2007 Jun 14;447(7146):799-816.
\bibitem{ErikLearnedMiller2003}
Erik Miller., "A new class of entropy estimators for multi-dimensional densities",.
{\em International Conference on Acoustics, Speech, and Signal Processing (ICASSP)}, 2003.
\bibitem{Darbellay-Vajda1999}
G. A. Darbellay and I. Vajda, "Estimation of the information by an
adaptive partitioning of the observation space," {\em IEEE Trans. on
Information Theory}, vol. 45, pp. 1315--1321, May 1999.
\bibitem{Geweke1982}
Geweke J., "The Measurement of Linear Dependence and Feedback
Between Multiple Time Series," {\em Journal of the American
Statistical Association}, 1982, 77, 304-324.  (With comments by E.
Parzen, D. A. Pierce, W. Wei, and A. Zellner, and rejoinder)
\bibitem{HastieTibshirani2002}
Hastie T, Tibshirani R, The Elements of Statistical Learning ,
Springer 2002.
\bibitem{chromatin-ChIP}
Heintzman ND, Stuart RK, Hon G, Fu Y, Ching CW, Hawkins RD, Barrera LO, Van
Calcar S, Qu C, Ching KA, Wang W, Weng Z, Green RD, Crawford GE, Ren B.,"Distinct and predictive chromatin signatures of transcriptional promoters and
enhancers in the human genome"., {\em Nat Genet. 2007} Mar;39(3):311-8.
\bibitem{PromFind}
Hutchinson GB., "The prediction of vertebrate promoter regions using
differential hexamer frequency analysis".,{\em Comput Appl Biosci}.
1996 Oct;12(5):391-8.
\bibitem{SignalProcMag2006}
Hudson, J.E., "Signal Processing Using Mutual Information", {\em
Signal Processing Magazine},Volume: 23,no: 6 pp:50-54, Nov. 2006.
\bibitem{John1989b}
H. Joe., ``Relative entropy measures of multivariate dependence".
{\em J. Am. Statist. Assoc}., 84:157–164, 1989.
\bibitem{Khandekar2004}
Khandekar M, Suzuki N, Lewton J, Yamamoto M, Engel JD., "Multiple,
distant Gata2 enhancers specify temporally and tissue-specific
patterning in the developing urogenital system".,{\em Mol Cell
Biol}. 2004 Dec;24(23):10263-76.
\bibitem{Kleinjan2005}
Kleinjan DA, van Heyningen V., "Long-range control of gene
expression: emerging mechanisms and disruption in disease"., {\em Am
J Hum Genet}. 2005 Jan;76(1):8-32.
\bibitem{Sanger-Histone}
Koch CM, Andrews RM, Flicek P, Dillon SC, Karaöz U, Clelland GK, Wilcox S, Beare
DM, Fowler JC, Couttet P, James KD, Lefebvre GC, Bruce AW, Dovey OM, Ellis PD,
Dhami P, Langford CF, Weng Z, Birney E, Carter NP, Vetrie D, Dunham I.,"The landscape of histone modifications across 1\% of the human genome in five
human cell lines".,{\em Genome Res}. 2007 Jun;17(6):691-707.
\bibitem{Kreiman2004}
Kreiman G., "Identification of sparsely distributed clusters of
cis-regulatory elements in sets of co-expressed genes".,{\em Nucleic
Acids Res}. 2004 May 20;32(9):2889-900.
\bibitem{CNSGata3}
Lakshmanan, G., K. H. Lieuw, K. C. Lim, Y. Gu, F. Grosveld, J. D.
Engel, and A. Karis. 1999. "Localization of distant urogenital
system-, central nervous system-, and endocardium-specific
transcriptional regulatory elements in the GATA-3 locus". {\em Mol.
Cell. Biol}. 19:1558-1568.
\bibitem{Shh}
Lettice LA, Heaney SJ, Purdie LA, Li L, de Beer P, Oostra BA, Goode D, Elgar G,
Hill RE, de Graaff E., "A long-range Shh enhancer regulates expression in the developing limb and fin and is associated with preaxial polydactyly".,{\em Hum Mol Genet}. 2003 Jul 15;12(14):1725-35.
\bibitem{Lieb2006}
Lieb JD, Beck S, Bulyk ML, Farnham P, Hattori N, Henikoff S, Liu XS,
Okumura K, Shiota K, Ushijima T, Greally JM., "Applying whole-genome
studies of epigenetic regulation to study human disease".,{\em
Cytogenet Genome Res}. 2006;114(1):1-15.
\bibitem{Mecp2}
Liu J, Francke U.,"Identification of cis-regulatory elements for MECP2 expression".,{\em Hum Mol Genet}. 2006 Jun 1;15(11):1769-82.
\bibitem{Fraenkel2006}
MacIsaac KD, Fraenkel E., "Practical strategies for discovering
regulatory DNA sequence motifs".,{\em PLoS Comput Biol}. 2006
Apr;2(4):e36.
\bibitem{Marko1973}
H. Marko, "The Bidirectional Communication Theory - A Generalization
of Information Theory", {\em IEEE Transactions on Communications},
Vol. COM-21, pp. 1345-1351, 1973.
\bibitem{Massey1990}
J. Massey, "Causality, feedback and directed information," {\em
Proc. 1990 Symp. Information Theory and Its Applications
(ISITA-90)}, Waikiki, HI, Nov. 1990, pp. 303–305.
\bibitem{Gata2}
Minegishi N, Ohta J, Yamagiwa H, Suzuki N, Kawauchi S, Zhou Y, Takahashi S,
Hayashi N, Engel JD, Yamamoto M., "The mouse GATA-2 gene is expressed in the para-aortic splanchnopleura and aorta-gonads and mesonephros region"., {\em Blood}. 1999 Jun 15;93(12):4196-207.
\bibitem{looping_scan_track}
Petrascheck M, Escher D, Mahmoudi T, Verrijzer CP, Schaffner W, Barberis A.,"DNA looping induced by a transcriptional enhancer in vivo"., {\em Nucleic Acids Res}. 2005 Jul 7;33(12):3743-50.
\bibitem{EnhancerBrowser}
Pennacchio, L. A., Ahituv, N., Moses, A., Prabhakar, S., Nobrega,
M., Shoukry, M., Minovitsky, A., Dubchak, I., Holt, A., Lewis, K.,
Plazer-Frick, I., Akiyama, J., DeVal, S., Afzal, V., Black, B.,
Couronne, O., Eisen, M., Visel, A., and Rubin, E.M. 2006., "In vivo
enhancer analysis of human conserved non-coding sequences", {\em
Nature}, 444(7118):499-502.
\bibitem{Enhancer_Prediction}
L.A. Pennacchio, G.G. Loots, M.A. Nobrega, and I. Ovcharenko, "Predicting
tissue-specific enhancers in the human genome", {\em Genome Research},
17(2), 201-11 (2007)
\bibitem{Silverman1997}
J. Ramsay, B. W. Silverman, Functional Data Analysis (Springer
Series in Statistics), Springer 1997.
\bibitem{CSB2007}
Rao A, Hero AO, States DJ, Engel JD, "Using Directed Information to build Biologically Relevant Influence Networks", {\em Proc. Computational Systems Bioinformatics (CSB)}, 2007.
\bibitem{Stoeckert2005}
Schug J., Schuller W-P., Kappen C., Salbaum J.M., Bucan M.,
Stoeckert C.J. Jr., "Promoter Features Related to Tissue Specificity
as Measured by Shannon Entropy"., {\em Genome Biology} 6(4): R33,
March 2005.
\bibitem{Segal2006}
Segal E, Fondufe-Mittendorf Y, Chen L, Thåström A, Field Y, Moore IK, Wang JP,
Widom J.,"A genomic code for nucleosome positioning".,{\em Nature}. 2006 Aug 17;442(7104):772-8.
\bibitem{Stuart2003}
Stuart RO, Bush KT, Nigam SK, "Changes in gene expression patterns
in the ureteric bud and metanephric mesenchyme in models of kidney
development", {\em Kidney International},64(6),1997-2008,December
2003.
\bibitem{DWE2005}
Sumazin P,  Chen G, Hata N ,  Smith A D., Zhang T, Zhang M Q., "DWE: Discriminant Word Enumerator", {\em Bioinformatics}, 21(1):31-38, 2005.
\bibitem{Cover1990}
Cover TM, Thomas JA, Elements of Information Theory, {\em Wiley-
Interscience}, 1991.
\bibitem{EnhancerBrowser2}
Visel A, Minovitsky S, Dubchak I, Pennacchio LA., "VISTA Enhancer
Browser--a database of tissue-specific human enhancers"., {\em
Nucleic Acids Res}. 2007 Jan;35(Database issue):D88-92.
















\end{thebibliography}
}
\end{document}